# Harnessing the polar vortex motion in oxide heterostructures

Pushpendra Gupta[1#], Mohit Tanwani[1#], Qi Xu[2#], Guanshihan Du[3#], Peiran Tong[2], Yongjun Wu[3], Zijian Hong[3,4,5] *, He Tian [2,4,6]*, Ramamoorthy Ramesh[7,8,9], Sujit Das[1]*

[1] Department of Materials Research Centre, Indian Institute of Science, Bangalore, 560012 Karnataka, India

[2] Center of Electron Microscopy, State Key Laboratory of Silicon and Advanced Semiconductor Materials, School of Materials Science and Engineering, Zhejiang University, Hangzhou 310027, China

[3] School of Materials Science and Engineering, Zhejiang University, Hangzhou 310027, China

[4] Institute of Fundamental and Transdisciplinary Research, Zhejiang University, Hangzhou 311121, China

[5] Zhejiang Key Laboratory of Advanced Solid State Energy Storage Technology and Applications, Taizhou Institute of Zhejiang University, Taizhou, Zhejiang 318000, China

[6] Pico Electron Microscopy Center, Hainan University, Haikou 570228, China

[7] Department of Materials Science and Engineering, University of California, Berkeley, Berkeley, CA, USA.

[8] Department of Physics, University of California, Berkeley, Berkeley, CA, USA.

[9] Materials Sciences Division, Lawrence Berkeley National Laboratory, Berkeley, CA, USA.

# P.G., M.T., Q.X., G.D. equally contributed

* Corresponding authors. hongzijian100@zju.edu.cn(Z.H.); hetian@zju.edu.cn(H.T.); sujitdas@iisc.ac.in(S.D.)

**Abstract**

Polar topology, an analogue of the magnetic topology, serves as a large playground for exotic physical phenomena with a wide range of multifunctional applications. Polar vortices and skyrmions are representative polar topologies that have been predicted to significantly enhance the functionality and information density of nanoelectronic devices due to their ultrasmall dimensions.[1,2] Despite these advantages, the practical realization of polar topologies in devices is impeded by the intrinsic challenges associated with their controlled motion and manipulation. Therefore, harnessing vortex manipulation—such as motion, on demand creation, annihilation, and shape transformation—is essential for practical device integration. However, vortex motion is often challenged by intrinsic physical limitations in collective lattice distortions and strong pinning effects from the surrounding environment, which remains elusive. In this study, we present real time observation of vortex motion in $PbTiO_3/SrTiO_3$ heterostructures, achieved through the application of localized pulsed electric fields and trailing bias fields from a conductive tip. Notably, the vortices exhibit reversible motion in response to the field direction. Furthermore, by precisely manoeuvring the conductive Atomic-Force-Microscopy tip along specific trajectories, we achieved controlled vortex reshaping, with reconfigured vortices showing remarkable stability over extended periods. This underline physical mechanism is further pinpointed by phase-field simulations, which revealed that the motion of the vortex boundary is controlled through the switching of the zigzag patterns of the vortex core. This study highlights the feasibility of harnessing vortex dynamics through external stimuli, advancing the fundamental physical understanding and prospects for next-generation polar vortex-based nanoelectronic devices.

**Polar vortex: A new pathway for exotic physical phenomena and enhanced information storage density.** The emergence of ferroelectric topologies such as polar vortices and skyrmions within ferroelectric-dielectric superlattices like $(PbTiO_3)_n/(SrTiO_3)_n$ (denoted as PTO/STO), makes a significant milestone for modern condensed matter physics and material science,[3,4,5,6,7,8] signalling the new era characterized by complex and emergent phenomena across oxide hetrostructures.[9,10] Polar topologies, described as regions of highly inhomogeneous spontaneous electric polarization arranged in swirling pattern within a few unit cells, exhibit intricate dynamics and properties that have captivated researchers and spurred investigations into their physical origin, functionality and potential technological applications.[1-2] One particular example is the polar vortex, which exhibits continuous polarization rotation surrounding a singularity-like vortex core.[11,12] A key advantage of polar vortices is their ability to process data at high frequencies (sub-THz signals) with energy efficiency.[13,14] Their nanoscopic dimensions (~10-15 nm) make them promising candidates for next-generation memory and logic devices, with an ultrahigh theoretical memory capacity (up to 60 Tb/inch$^2$).[15] This could potentially revolutionize information storage and computing technology with remarkably low energy consumption if these entities can be manipulated in a controlled manner, which could largely contribute to the recent development of internet-of-things, artificial intelligence and large language models.[16] In the context of magnetic memory systems, the collective motion of vortex and skyrmion has been achieved through the spin–transfer torque or spin–orbit torque, which can be utilized to design racetrack memory for data storage and logic applications.[17] Meanwhile, the real-time observation of polar vortex or vortex boundary motion under an applied electric field has yet to be realized due to both, the lack of the corresponding interactions in ferroelectrics and the strong coupling of polar vortices to lattice, causing underlying dynamics of such polar vortices largely unexplored. From a theoretical perspective, the collective motion of the polar vortices is significantly hindered by their intrinsic sluggish motion dynamics. Their movement involves complex interactions with lattice distortions, defects, and external fields, rendering them highly sensitive and highly pinned to their surroundings[18,19] which complicates the precise control and manipulation, thereby impeding efforts to realize their practical uses.[20] Recent studies have demonstrated that structural defects such as dislocations, can be actively manipulated using external electric fields, e.g., dislocations in single-crystalline zinc sulphide (ZnS) can move back and forth in response to the direction of the electric field.[21] One natural question to ask: can we use an electric field to trigger the motion of the topological defects such as polar vortices? Harnessing collective behaviour of polar vortices may enable deterministic nanoscale control and device functionalities, as well as revel previously inaccessible vortex interactions and dynamic phenomenon.[22] Our study seek to push the boundaries of physical understanding of this field by leveraging insights from dislocation dynamics and translating them in the context of polar vortex behaviour. In this study, we investigate the interplay of electrostatic energy landscape by applying external bias field to the PbTiO$_3$/SrTiO$_3$ heterostructures, and revealing the underlying collective dynamics of the polar vortices. It is revealed that an external bias can harness the motion, annihilation, creation, and morphological modification of polar vortices. The movement and morphological changes of the vortices are characterized using advanced microscopy techniques, and the insights into the underlying physical mechanisms are provided by phase field simulations. This study offers a deep understanding of the fundamental physics and pathway to harness the unique properties of polar vortices, which could be employed in high-density, low-power non-volatile memory and reconfigurable nanoelectronics, where robust, dynamic control of vortices facilitates ultra-compact data storage and adaptive logic devices.

**Growth of PTO/STO heterostructures and stabilization of polar vortices.** 20 uc STO/20 uc PTO/20 uc STO trilayer and [PTO$_{16}$/STO$_{12}$]$_{10}$ superlattices were synthesized on 5 nm buffered SrRuO$_3$ on single-crystalline DyScO$_3$ (001)$_{pc}$ substrates using reflection high-energy electron diffraction (RHEED)-assisted pulsed-laser deposition (Supplementary Fig. 1, details in Methods). The X-ray diffraction measurements around the (002)$_{pc}$ reflection (where, "pc" denotes pseudo-cubic) confirmed the crystalline quality of the deposited films. X-ray diffraction around the (002)$_{pc}$-reflection of the superlattice revealed a distinct double peak, indicative of the coexistence of vortex phase (labelled V) and ferroelectric phase with in-plane polarization (labelled FE) (Supplementary Fig. 2a), consistent with previous findings.[23] The V$_{(0)}$ phase and FE$_{(0)}$ phase arise from the different out-of-plane effective lattice parameters of the vortex (c= 4.05 Å) and ferroelectric $a_1/a_2$ ($c_{FE}$ = 3.959Å) phases, respectively. Reciprocal space mapping (RSM) around the (002)$_{pc}$-diffraction condition for both the trilayer (Supplementary Fig. 2b) and superlattice (Supplementary Fig. 2c) highlights the presence of side lobes extending in both in-plane and out-of-plane directions (marked by arrows). These side lobes indicates the presence of well-ordered vortices within the films,[23] with in-plane periodicity of ~14 nm and ~10 nm for the trilayer and superlattice, respectively (details in Methods). The plane-view high-angle annular dark-field scanning transmission microscopy (HAADF-STEM) for superlattice images provided further evidence of the existence of vortices oriented along the [100]$_{pc}$ and [010]$_{pc}$ crystallographic directions (Fig. 1a, left panel, details in Methods). The zoomed regions of these images, displayed in the right panel, offer a more detailed view of these features. Notably, $a_1/a_2$ superdomains were absent in the TEM images, likely due to alterations in boundary conditions introduced during TEM sample preparation (details in Methods), as pointed out previously.[1] The cross-sectional HAADF and dark field (DF) images of the trilayer and superlattice confirm the presence of vortices within the trilayer and superlattice structure (Supplementary Fig. 3a-b). Additionally, cross-sectional HAADF and bright field (BF) STEM images of trilayer also validate the presence of vortices in the trilayer samples (Fig.1b and Supplementary Fig. 3c-d). These findings demonstrate the emergence of nanoscale vortices, consistent with the previous reports.[1,23]

The presence and arrangement of vortices in the trilayer and superlattice samples, were further investigated using piezoresponse force microscopy (PFM) (Fig. 1c-d; detailed in Methods). PFM scans of the trilayer in lateral PFM (LPFM) amplitude, along with the corresponding phase images, distinctly reveal pure vortices, as highlighted by white dashed lines (Fig. 1c). However, in the case of the superlattice, super-domains of $a_1/a_2$ (highlighted by blue dashed lines) were also observed along with the vortices (highlighted by white dashed lines) (Fig. 1d), which aligns with the HRTEM observations. To get the physical insight into the switching dynamics of polar vortices and their response to an external electric field, Switching Spectroscopy PFM (SS-PFM) measurements were performed (Fig. 1e–f; details in Methods). SS-PFM enabled the capture of the spatial distribution of vortex switching behaviour, revealing the local energy landscape associated with polarization reversal. Notably, region with reduced coercive field were identified along vortex boundaries, playing important role in harnessing of vortex motion. The measurements were conducted over a dense grid of 50×50 points within a 2×2 μm² area. For the trilayer sample (Fig. 1e), the spatial map of the positive nucleation bias ($V_{C+}$, left panel) reveals well-defined domains with opposite polarization (highlighted by cyan dashed boundaries). The corresponding map of the average coercive bias ($V_C$, right panel) for the same area indicates a locally reduced coercive field near the vortex boundaries. A similar trend is observed in the superlattice sample (Fig. 1f), where the coercive voltage is also significantly reduced at the vortex boundaries. To analyze this observation further, we extracted and averaged hysteresis loops from 10 representative points located both within the vortex cores and in adjacent vortex boundary regions (Fig. 1g) for both trilayer (top) and superlattice (bottom) samples. The averaged loops confirm a consistent reduction in the coercive field at the vortex boundaries, indicating that these regions are more susceptible to trailing electric fields. This highlights the potential to harness vortex boundary motion using low electric field bias, and the critical role that vortex boundaries play in the nucleation and switching of the polar vortex phase.

The underline physical mechanism supporting the PFM observations are further revealed by phase-field simulations, which confirm the presence of a pure vortex phase in the trilayer (Fig. 1g) and a mixed phase of vortices with $a_1/a_2$ twin domains in the superlattice (Fig. 1i; details in Methods). These distinct topological configurations arise from a delicate interplay among Landau, elastic, electric and gradient energies contributions, reflecting the intricate energy landscape that governs the systems' equilibrium staes.[1] The three-dimensional polarization structure of the vortices reveals a swirling polarization, characterized by a gradual rotational change in the polarization phase (right panels of Fig. 1g–i). The low amplitude of these vortices results in their distinct dark contrast in phase images, allowing for clear visualization (Fig.1c-d). Having confirmed the presence of the vortex phase in both the trilayer and superlattice, a critical next step involves harnessing their movement, temporal stability under external stimuli. Such insights are essential for elucidating the mechanisms by which these vortices can be manipulated and potentially harnessed for functional device applications.

**Harnessing the vortex motion under a trailing bias field.** Vortices are complex topological entities, whose manipulation requires a nuanced interplay of various energies. These vortices are distinguished by their boundaries, which, upon moved, prompt the vortex boundary to expand or contract accordingly. To control this behavior, a trailing bias field was applied along a lithographically defined channel (indicated by the cyan arrow; details in Methods), enabling directed harnessing of vortex motion from left to right. Reversing the field then restored the vortices to their original positions (Fig. 2-3, Supplementary Fig. 4-7). The pristine LPFM scan of the trilayer shows pure vortices with distinct boundaries, evident as dark contrast in the PFM phase (Fig. 2a). A positive trailing bias field of +5 V was applied for one minute through the lithographically defined channel (CHP1), to the pristine state (Fig. 2a), locally modifying the electrostatic energy landscape. This reduces the energy barrier at the vortex boundary, thereby initiating the harnessing of vortex motion (Supplementary Fig. 8). After removing the trailing field, a subsequent scan showed that the vortex boundary (shown by dashed white line) had shifted to the right in response to the applied field (CHP1) (Fig. 2b, indicated by the bold blue arrow). After successive applications of the third (CHP3), fifth (CHP5), and sixth (CHP6) trailing fields, each under +5 V for one minute, the vortex boundary progressively shifted further to the right, with a cumulative displacement of ~1170 nm observed after CHP6 (Fig. 2c-e). The complete details of all positive trailing field steps are provided in Supplementary

Fig. 3. This movement under the positive trailing bias field demonstrates the controlled movement of the vortex boundary, effectively enabling the harnessing of the vortex boundary expansion.

To reverse this phenomenon, the polarity of the trailing bias field was switched while the magnitude remained constant (e.g., under -5 V). Using the state after previous motion (Fig. 2e) as a starting point, a negative trailing field was applied through the channel, pushing the vortex boundary to the leftward side (highlighted by bold blue arrow). Sequential application of several negative trailing fields (i.e., CHN1, CHN3, CHN5, CHN6), each for one minute, progressively shifted the vortex boundary from right to left again, resulting in a total displacement of ~890 nm after CHN5 (Fig. 2f-h; each step necessary to change the boundary is shown in Supplementary Fig. 5), which effectively move the boundary to its original position (Fig. 2i). These observations highlight that the polarity reversal of electric field can enable deterministic control of the directional vortex boundary motion. Moreover, *in-situ* STEM experiments were conducted to investigate the harnessing of the vortex motion (Fig. 2j, details in Methods). In the initial cross-sectional trilayer STEM image, as shown in Fig. 2j(I), the vortices exhibit a zig-zag alignment. Upon applying an electric field (+17 V) through the tungsten tip, as shown by arrow in Fig. 2j(II), the vortex cores near the tip begin to align, resulting in the formation of a distinct boundary (marked with an orange dashed line) between the zig-zag arranged vortices. Interestingly, after the removal of the electric field, the vortices did not revert to their original positions but instead demonstrated stability in their new configurations. Subsequent application of a +17 V electric field further shifted the vortex boundary toward the right, as observed in Fig. 2j(III). These findings are consistent with the PFM data presented earlier, reinforcing the reproducibility and reliability of the harnessing of vortex dynamics.

Our experimental findings are further corroborated by phase field simulations, providing a deeper physical understanding of the observed phenomena (Fig. 2k-n, Supplementary 8). The initial simulated configuration reveals the presence of vortices, evident in both the plane-view (Fig. 2k, top panel) and cross-sectional (Fig. 2k, bottom panel) representations. Upon applying a positive electric field through a designated channel (indicated by black dashed lines), a shift in the vortex boundary is observed (Fig. 2l). This movement, clearly visible in both plane-view and cross-sectional simulations, demonstrates the boundary's displacement toward the right. Continued application of the positive electric field causes a progressive shift of the vortex boundary in the same direction, consistent with the experimental results. After removing the applied field, the vortex boundaries exhibited stable behavior, maintaining their displaced positions. To investigate the movement of the vortex boundary in the opposite direction (from right to left), a negative electric field was applied. As shown in Fig. 2n-p, the vortex boundary, marked by black dashed lines, shifted progressively toward the left. This observation demonstrates that the vortex boundary can be manipulated horizontally by applying a trailing bias field and can be restored to its original position by reversing the polarity of the field. The energy release associated with vortex motion is illustrated in Supplementary Fig.8. The energy profile reveals a peak during the transition as the vortices shift to a new position, followed by a reduction to a minimum during the subsequent relaxation phase. This indicate that the injection of the electric energy could help the system to overcome the energy barrier for the switching between two energy equivalent states. Interestingly, the energy attains a maximum when the electric field direction switches from negative to positive. This phenomenon can be attributed to the polarization direction being antiparallel to the applied field during the transition, leading to an energy penalty before the system stabilizes. These findings highlight the dynamic energy landscape associated with vortex manipulation under external stimuli.

To further explore the harnessing of vortex dynamics in the superlattice, a series of PFM experiments were performed (Fig. 3, Supplementary Figs. 6-7). The pristine LPFM images, displaying both amplitude (left) and phase (right) contrasts, reveal the nanoscale features in detail (Fig. 3a). The amplitude image clearly shows vortices (dark contrasts, marked by white dashed lines) coexisting with ferroelectric $a_1/a_2$ superdomains (marked by short blue dashed lines). A positive trailing field (+5 V for one minute) was applied through a channel (CHP1, indicated by the cyan arrow), and scans conducted after the field is removed revealed the formation of a distinct vortex boundary (Fig. 3b). Sequential bias applications through additional channels (e.g., CHP5, CHP8) caused the vortex boundary to shift progressively to the right (Fig. 3b-d, highlighted by bold blue arrow; Supplementary Fig. 6). Conversely, applying a negative trailing field (e.g., CHN1, CHN4) induced a leftward boundary shift, with disruptions observed after several field cycles (Fig. 3e-f, Supplementary Fig. 7). These experiments demonstrate the harnessing of vortex boundaries in a controlled fashion using trailing fields, enabling their movement from left to right and their subsequent restoration to initial positions via reverse bias fields. To validate this observation further, *in-situ* biased STEM experiments were conducted on a plane-view superlattice sample (Fig. 3g; details in Methods and Supplementary Fig. 9). The initial STEM image reveals the boundary vortices oriented along $[010]_{pc}$ and $[100]_{pc}$ directions, highlighted by white and yellow dashed lines, as shown in Fig. 3g (I). The white dashed line serves as a reference, and the red dot marks the position where a +20V bias electric field was applied using a tungsten tip. Upon applying the bias field, the vortex boundary (yellow dashed line) moved closer to the reference line in Fig. 3g(II). With the tungsten tip moving forward, the boundary continued to move closer to the reference line, as depicted in Fig. 3g (III), eventually resulting in the alignment of all vortex cores in a single orientation (Fig. 3g(IV)). This indicates that the vortex boundary moves via asymmetric growth and suppression of adjacent vortex domains, driven via local energy minimization under applied bias. Phase-field simulations further confirm this observation. Starting with a mixed phase of vortices and coexisting $a_1/a_2$ twins in the superlattice system (Fig. 3h(I), details in Methods), an electric field is applied through a simulated conducting tip modelled by a Lorentz distribution. The vortices extended along the direction of tip movement, demonstrating the harnessing of vortex movement under bias (Fig. 3h(II-IV)). After the movement of vortices, stability of the final achieved state was observed over time (Supplementary Fig. 7e-f). These results confirm that the achieved vortex states are exceptionally stable, highlighting their potential for applications in reliable memory devices.

**Stability in harnessing vortex motion.** The stability of newly formed vortex boundaries is crucial for ensuring the reliability of memory and logic devices. Understanding the long-term behaviour of these vortex boundaries is essential for consistent device performance. To address this, a detailed investigation of their time-dependent evolution was conducted to show their stability and robustness (Fig. 4, Supplementary Fig. 10). To validate the reproducibility and stability of vortex boundary motion, a new region on the trilayer sample was selected, and the procedures described earlier (Fig. 2) were replicated. Pristine LPFM scans confirmed the presence of well-defined vortices in the region (Fig. 4a). Controlled movement of the vortex boundary was achieved by applying a trailing bias through a lithographically defined channel (Fig. 4b). Follow-up scans conducted after 2, 8, 30, and 120 days demonstrated the robust stability of the harnessed vortex boundary, confirming its long-term retention (Fig. 4c-f). The stability of vortex boundary motion was further validated using *in-situ* STEM measurements (Fig. 4g). The initial STEM image (Fig. 4g(I)) reveals two orientations vortices, with their boundary marked by a yellow dashed line. A +15V electric field was applied at the location indicated by the red point using a tungsten tip (Fig. 4g(II)). The application of the field caused the vortex boundary to shift, moving in tandem with the tungsten tip (Fig. 4g(III)). Upon removal of the electric field, the vortex boundary exhibited remarkable stability, with no discernible changes observed even after 20 days, which is consistent with PFM experiment (Fig. 4g(IV)). The results indicate that the vortex boundary remains stable without significant variation over an extended period, affirming its temporal stability until external stimuli are applied to alter its state. This investigation provides valuable insights into the durability and consistency of vortex motion, contributing to the development of robust vortex-based devices.

**Stochastic nature of vortices and creation-annihilation of vortex ring, pair of rings under pulsed field.** Following the harnessing of vortex motion through trailing fields, the stochastic behavior of vortices—including vortex creation, annihilation, and the formation of new vortex rings, pair of rings—was observed upon the application of a pulsed electric field through the PFM tip at a defined location in the superlattice (Fig. 5a, marked by a solid yellow circle). Subsequent scans conducted after the field removal revealed these dynamic transformations in vortex configurations (Fig. 5 and Supplementary Fig. 11; detailed in Methods). To further investigate the stochastic nature of vortex behavior, two distinct vortices were identified within the superlattice: vortex 1, positioned above the solid yellow circle (located at position 1), and vortex 2, below the circle (located at position 2) (Fig. 5a). After applying an electric pulse, both vortices displayed predominantly random trajectories, underscoring the non-deterministic nature of their motion. Notably, the adjacent ferroelectric $a_1/a_2$ superdomains remained unaffected by the applied pulsed field. An initial 1V pulsed field applied for one minute produced minimal alterations in vortex movement upon rescanning the same region after field removal (Supplementary Fig. 11a-c). Incremental increases in the applied voltage, beginning at 1V and increasing in 0.5V steps for one minute each, revealed significant transformations at 2V (Fig. 5b, Supplementary Fig. 11d). These transformations included the emergence of stochastic nature of the vortices, new vortices, morphological changes (highlighted by white dashed lines), annihilation of certain vortices, and the formation of a vortex ring (denoted by a yellow dashed line), pair of ring (denoted by a green dashed line).

Increasing the magnitude of the applied pulsed field induced pronounced transformations within the vortex domains, leading to the disappearance of the vortex ring, pair of rings (Fig. 5c-d, Supplementary Fig. 11e-i). Notably, the vortex ring and pair of rings reappeared upon applying +4V and +5V pulses for one minute each at the same location (Fig. 5e-g, Supplementary Fig. 10j-k). At higher fields, the ring vanished again (Fig. 5h, Supplementary Fig. 10l), only to reappear under 6.5V (Fig. 5i, Supplementary Fig. 11m-o). These observations highlight the stochastic nature of the vortices and the sensitivity of vortex dynamics under applied pulsed fields (Fig. 5m-n). Applying a maximum +7V pulse and extending the pulse duration to two and three minutes (Fig. 5j-l, Supplementary Fig. 11o-r) stabilized the vortex structures, resulting in persistent vortices. When the field was reduced to 0V, further stochastic vortex behaviors were observed (Supplementary Fig. 12). Incrementally increasing the negative pulsed field from -1V to -7V produced distinct vortex recurrence patterns, indicating domain switching events triggered by the electric field (Supplementary Fig. 12). These events nucleated polarization structures with unique morphologies. These findings emphasize the high sensitivity of vortices to electric fields and underscore the critical role of field magnitude and duration in controlling switching events for precise management of polar vortex dynamics. The formation and behavior of vortex rings were further validated through in-situ STEM imaging (Fig. 5o-p). Notably, in certain regions, paired vortex rings were also observed (Fig. 5q), providing additional insight into the complex and dynamic nature of vortex structures under applied fields. The origin of formation of hidden vortex rings can be attributed to the excitation from charge injection by the PFM tip, which can affect the local strain gradients, field gradient, broken symmetry, or boundary constraints. Additionally, when a field was applied through a lithographic channel (orange line, Supplementary Fig. 13) during simultaneous scanning, the vortex boundaries responded asymmetrically to field polarity (details in Methods). A +5V field caused partial melting of vortex boundaries on the left, while a -5V bias led to partial melting on the right. This asymmetry arises from the interplay between electrostatic forces and vortex dynamics: positive fields induce repulsion, destabilizing vortices on the left, while negative fields create attraction, moving the vortices to the right. These results demonstrate the tunability of vortex boundaries using electric fields, offering insights for optimizing vortex properties in applications.

To further improve device tunability, precise control over vortex boundaries in defined forms and at specific locations is essential. Applying trailing voltages of ±5 V through lithographically specified channels allowed for precise on-demand control of vortex boundaries and moulding them into certain patterns such as lines, spirals, and stars (Supplementary Fig. 14-16). Upon removal of the field, LPFM scans consistently showed changes in phase and boundary movement in both trilayer and superlattice samples. These results were confirmed by phase-field simulations, which showed that vortex boundaries can be robustly tuned by playing the local energy under applied bias.

**Conclusion.** This work first time demonstrates the harnessing of controlled manipulation of vortex motion in trilayer and superlattice oxide systems through various electric field applications, including trailing and pulsed bias fields. Notably, vortex boundaries were moved over distances exceeding 1100 nm using a trailing bias field, and reversing the bias successfully returned the boundaries to their initial positions. These moved vortices exhibited remarkable temporal stability, highlighting their robustness. These findings unveil new avenues for harnessing topological polar structures in nanoscale electronic systems, where the ability to precisely manipulate vortex motion and configuration opens the door to functional applications such as ferroelectric racetrack memories and reconfigurable logic devices. Crucially, the application of localized electric fields not only drives vortex transformations at targeted sites but also induces long-range modifications across hundreds of nanometers, reflecting the inherently collective and nonlocal nature of the underlying order-parameter dynamics. Furthermore, the emergence of hidden topological phases—such as vortex rings and ring pairs—in systems exhibiting coexistence of vortices and $a_1/a_2$ twins suggests rich energy landscapes that can be engineered through subtle field-mediated interactions. Looking ahead, deterministic control over these topological excitations holds promise for realizing next-generation nanoelectronic components, including ultrasmall, field-tunable capacitors and transistors, pushing the boundaries of energy-efficient, high-density device architectures.

**Acknowledgement**

S.D. acknowledges Science and Engineering Research Board, Scheme for Transformational and Advanced Research in Sciences (MoE-STARS/STARS-2/2023-0048) and S.D. thanks Indian Institute of Science start up grant for financial support. P.G. acknowledge to Indian Institute of Science for Institute of Eminence (IoE) fellowship. The financial supports from Natural Science Foundation of Zhejiang Province (LR25E020003, ZH; LD24E020003, YW), the National Natural Science Foundation of China (No. 92166104, No. 92463306, ZH; No. U21A2067, YW & HT; No. 12125407, HT) are acknowledged. ZH further acknowledges the Fundamental Research Funds for the Central Universities (2023QZJH13) and a start-up grant from Zhejiang University. The phase-field simulation was performed with the Mu-PRO software package (https://muprosoftware.com). ZH are grateful for the technical support by Nano-X from Suzhou Institute of Nano-Tech and Nano-Bionics, Chinese Academy of Sciences (SINANO). H. T. acknowledges the National Key Research and Development Program of China (No. 2021YFA1500800).

**Methods**

**Growth of the trilayer and superlattice.** 20 uc PbTiO$_3$/20 uc SrTiO$_3$/20 uc PbTiO$_3$ (PTO/STO/PTO) trilayer and [16 uc PbTiO$_3$/12 uc SrTiO$_3$]$_{10}$ superlattices (total 100nm) and were deposited on top of a 5 nm SrRuO3 film grown on single-crystalline DyScO$_3$ (001)$_{pc}$ substrates using reflection high-energy electron-diffraction-assisted (RHEED) pulsed laser deposition equipped with KrF excimer laser ($\lambda$=248 nm). Ceramic targets of SrRuO$_3$, PbTiO$_3$ and single crystal SrTiO$_3$ target were used for the SRO, PTO and STO layer deposition, respectively. SrRuO$_3$ film was grown at 700°C with 100 mTorr oxygen pressure. Further substrate was cooled down to 600°C for the deposition of PTO and STO layers and the oxygen pressure was maintained to 50 mTorr. The laser intensity was 0.9 J/cm2 with a repetition rate of 10 Hz for all the films. RHEED was used during deposition to observe a layer-by-layer growth mode. RHEED oscillation for the trilayer and superlattice films have been shown in the Supplementary Fig.1a-d, where each oscillation represents the one-unit cell. The clear oscillations have been observed during the growth of all layers, which indicate the 2D growth of the samples. After deposition, the thin films were annealed for 10 minutes under 50 Torr oxygen pressure to promote full oxidation and then cooled to ambient temperature in the same pressure.

**X-ray diffraction.** Structural characterization of the PTO/STO/PTO trilayer and PTO/STO superlattices were carried out using a D8 Brucker X-ray diffractometer using Cu-K$\alpha$ radiation ($\lambda$ = 1.5405 Å). The high crystalline quality of the films, and the smooth nature of the interfaces, was confirmed from $\theta$-$2\theta$ symmetric XRD scans around the 002pc reflection, which show strong superlattice peaks and mixture of vortex and ferroelectric phase.

**Reciprocal space mapping.** In order to obtain a comprehensive picture of the vortices of the superlattices, as well as information on the in-plane and out-of-plane ordering vortex structures, further structural characterization was done using reciprocal space mapping around (002)$_{pc}$ diffraction condition. The diffraction allows to detect the weak diffracted intensities arising from the lattice modulations associated with the polar-vortex present in the trilayer and the superlattice.

**Conventional and scanning transmission electron microscopy.** The cross-sectional samples of trilayer and plane-view samples of superlattice were cut into lamellas with the widest faces perpendicular to the [010]$_{pc}$ and [001]$_{pc}$ direction, respectively, using Focused Ion Beam (FEI Quanta 3D FEG) for observation by transmission electron microscopy. We used spherical aberration-corrected electron microscopy (FEI Titan G2 80-200 ChemiSTEM, 30 mrad convergence angle, 0.8 Å spatial resolution) to acquire STEM images. The boundary conditions may be slightly altered during sample preparation due to the thinning process. However, the vortices remain clearly observable. The STEM images were recorded with 2048 × 2048 pixels for each frame and a dwell time of 1 μs per pixel. In consideration of the potential slight damage caused by ion beam in FIB, we minished parameters including voltage and electric current of ion beam down to 2 kV/27 pA to minimize the negative and unnecessary surface damage.

**Piezo-force Microscopy.** To study the vortices motion piezoresponce force microscopy (PFM) (Asylum Research MFP-3D origin+) have been employed using Budget sensor Platinum/chromium coated probes (Multi75G, spring constant ≈3N/m, resonance frequency 75KHz, tip radius 25 nm). We used the Dual AC resonance-tracking (DART) mode for acquiring PFM images. For acquiring lateral PFM signal tip was tuned to respective contact resonance frequency (~670KHz). . The drive amplitude has been kept to 0.6V to 1.25 V for all scans.

**Switching spectroscopy piezo-force microscopy.** To investigate the local polarization switching behaviour and spatial distribution of nucleation bias, Switching Spectroscopy piezoresponse Force Microscopy (SS-PFM) measurements were performed on both trilayer and superlattice samples. The measurements were conducted over a dense grid of 50×50 points within a 2×2 μm² area, allowing for high-resolution mapping of switching characteristics. Two hysteresis loops were acquired at each grid point with drive amplitude of 650mV using a triangular voltage waveform. Voltage was swept from 0 V to a positive maximum, then to a negative maximum, and finally back to 0 V. For the trilayer samples, the max voltage magnitude applied was 5V, while for the superlattice samples max voltage magnitude 8 V was applied.

**Voltage controlled piezo-force microscopy for applying trailing bias field.** For the movement of vortex boundary trailing bias field has been employed through the channel. In each channel and the moving tip having ±5 V bias voltage with average velocity of ~60 nm/sec. After the application of bias voltage, area has been scanned using DART mode with 0.75 to 1.25V drive amplitude. For moving the boundary, multiple channels have been employed (one at a time). For creating different shape of vortex boundary, the channel of different shape has been used.

**Voltage controlled piezo-force microscopy.** Amplitude (0 to 7 V) in steps of 0.5 V was applied at the centre of the designated area for a set time (one minute, unless otherwise stated). During the scan, the applied pulsed voltage was withdrawn, and the scan was performed with a 1.25 V drive amplitude. As previously noted, a further bias voltage was provided in decrement of 7 V to 0 and then 0 to -7 V. After applying bias voltage for 1 minute, the bias voltage was withdrawn and a scan was performed.

**Voltage controlled time evolution through Piezo-force Microscopy.** For the time evolution measurement, a constant bias voltage of +5V was applied for the different time duration. After that, the bias voltage removed and scan was performed. Throughout the trials, PFM scans were performed to capture the response of the polar vortices to the applied electric field.

*In-situ* **biasing TEM**. The in-situ biasing was performed using a Pico-Femto TEM electrical holder from Zeptools, controlled by an STM probe control system. An electrical bias was applied between a tungsten tip, which acts as a mobile electrode. The SRO conducting layer, which acts as the other electrode is connected to the holder ground. The input voltage is applied between the sharp conductive tip and the sample. During the whole *in-situ* experiments, we blocked the incident electron beam during all unnecessary moments, which helped to minimize the potential impacts of electron-beam irradiation.

**Phase field Simulation.** The spontaneous polarization vector ($\vec{P}$, i = 1, 3) was chosen as the order parameter. The polarization dynamics were determined by resolving the time-dependent Ginzburg-Landau equation[24,25]:

$$\frac{d\vec{P}}{dt} = -L\frac{\delta F(\vec{P})}{\delta \vec{P}}$$

Where $\vec{P}$ is the spontaneous polarization vector, and t is the evolution time step. The kinetic coefficient L is related to the domain wall mobility. F is the free energy functional of the system, consisting of four contributions (the Landau, mechanical, electric, and gradient energy densities), integrated over the entire volume (V) of the film:

$$F(\vec{P}) = \int (f_{elas} + f_{elec} + f_{grad} + f_{land})\, dV$$

Details of the free energies and the simulation parameters can be found in previous literature.[1,24,26,27]

The simulation setup differs between the trilayer and superlattice samples in terms of layer configuration and grid dimensions. Both systems are modeled using a three-dimensional grid, where each point represents a unit cell. Periodic boundary conditions are applied in the in-plane directions, while a superposition method is used for the out-of-plane direction. For the trilayer sample, the structure consists of 20 unit cells (uc) of $SrTiO_3$ (STO), 20 uc of $PbTiO_3$ (PTO), and another 20 uc of STO, all grown on a $DyScO_3$ (DSO) substrate. The simulation grid is 400×200×120, which includes 30 grid layers for the substrate, 60 layers for the trilayer film, and the remaining layers for air. In the superlattice sample, alternating stacks of [16 uc of PTO

/ 12 uc of STO] are repeated to reach a total thickness of 100 nm, deposited on a 5 nm SrRuO$_3$ buffer layer on a DyScO$_3$ substrate. The simulation grid for this setup is 200×200×250, which includes 30 grid layers for the substrate, 190 layers for the superlattice film, and the remaining layers for air. An iterative perturbation method is used to consider the elastic anisotropy for the PTO and STO layers.[28] The normalized time step is set as 0.01 in this study.

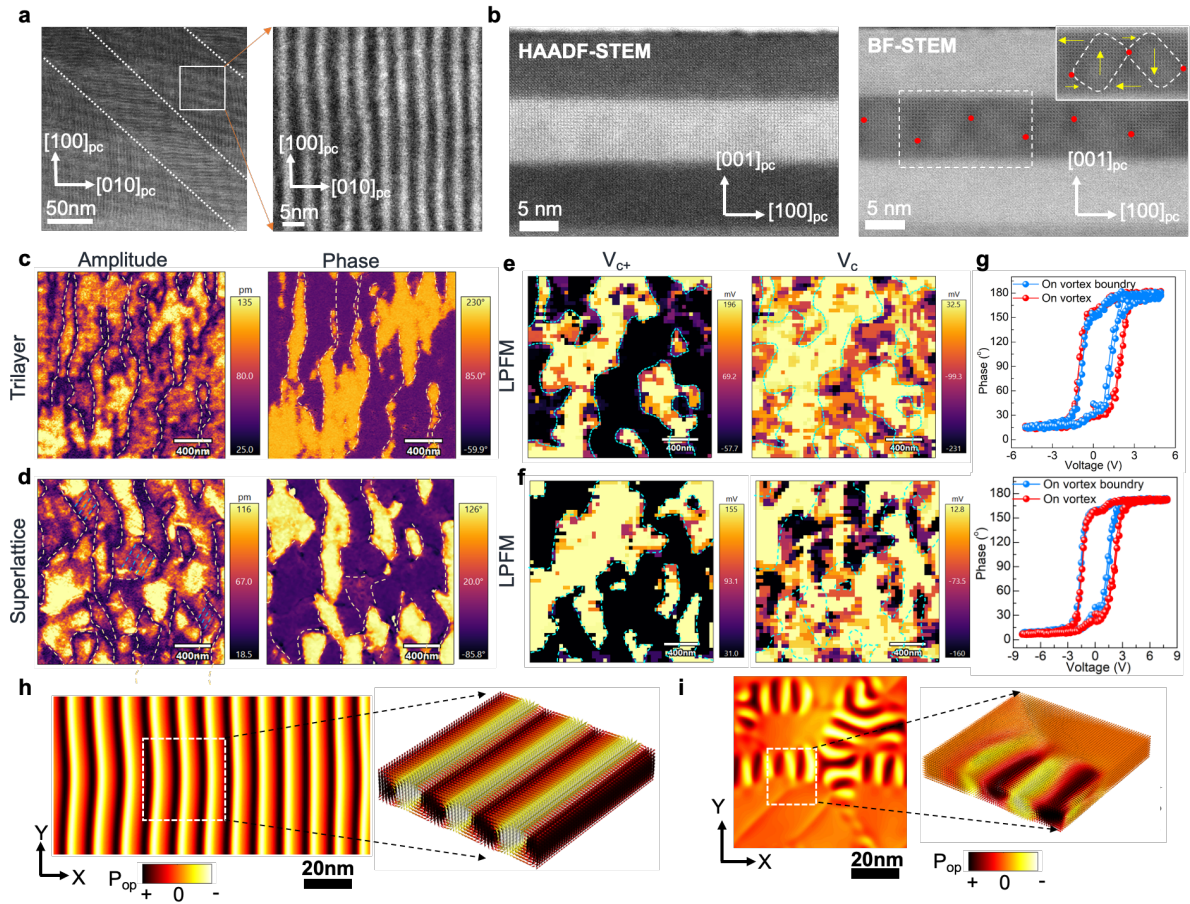

**Figure 1| Structure and stabilization of polar vortices in STO/PTO/STO trilayer and PTO/STO superlattices grown on DyScO₃ (001)$_{pc}$ substrates. a**, DF-TEM images in the plane view of the superlattice reveal the vortices are oriented along the [100]$_{pc}$ and [001]$_{pc}$ crystallographic directions. **b,** Cross-sectional HAADF STEM (left panel) and bright field (BF) STEM (right panel) images for the trilayer samples show contrast variations within the PTO layers corresponding to polar structures. Yellow arrows indicate the local polar directions, while red dots denote the vortex cores. **c-d,** LPFM scan for the trilayer (c) and for the superlattice (d) with amplitude (left) and phase (right). These scans illustrate the presence of a pure vortex phase (white dashed lines) in the trilayer sample, while the superlattice exhibits mixture of vortices (white dashed lines) with $a_1/a_2$ superdomains (blue dashed lines). The dark contrast in the amplitude represents the boundary of the vortex. **e-f,** SS-PFM mapping for trilayer (e) and superlattice (f) reveals spatial distribution of switching behaviour for vortex. The positive nucleation bias ($V_{c+}$, left panel) showing domains with different polarity having different contrast (marked by cyan dashed line) and the average coercive bias ($V_c$, right panel)

of same region shows vortex boundary having significantly lower coercive field. **g,** represents the hysteresis loops extracted from the vortex and near vortex boundary for trilayer (top) and superlattice (bottom) illustrating vortex boundary having lower coercive bias compared to within the vortex demonstrating vortex boundary motion can be harnessed using lower electric field. **g-i**, The phase field simulated images for the trilayer and superlattice, respectively. The zoomed images have been shown in the right side of the image 3D view of the vortices.

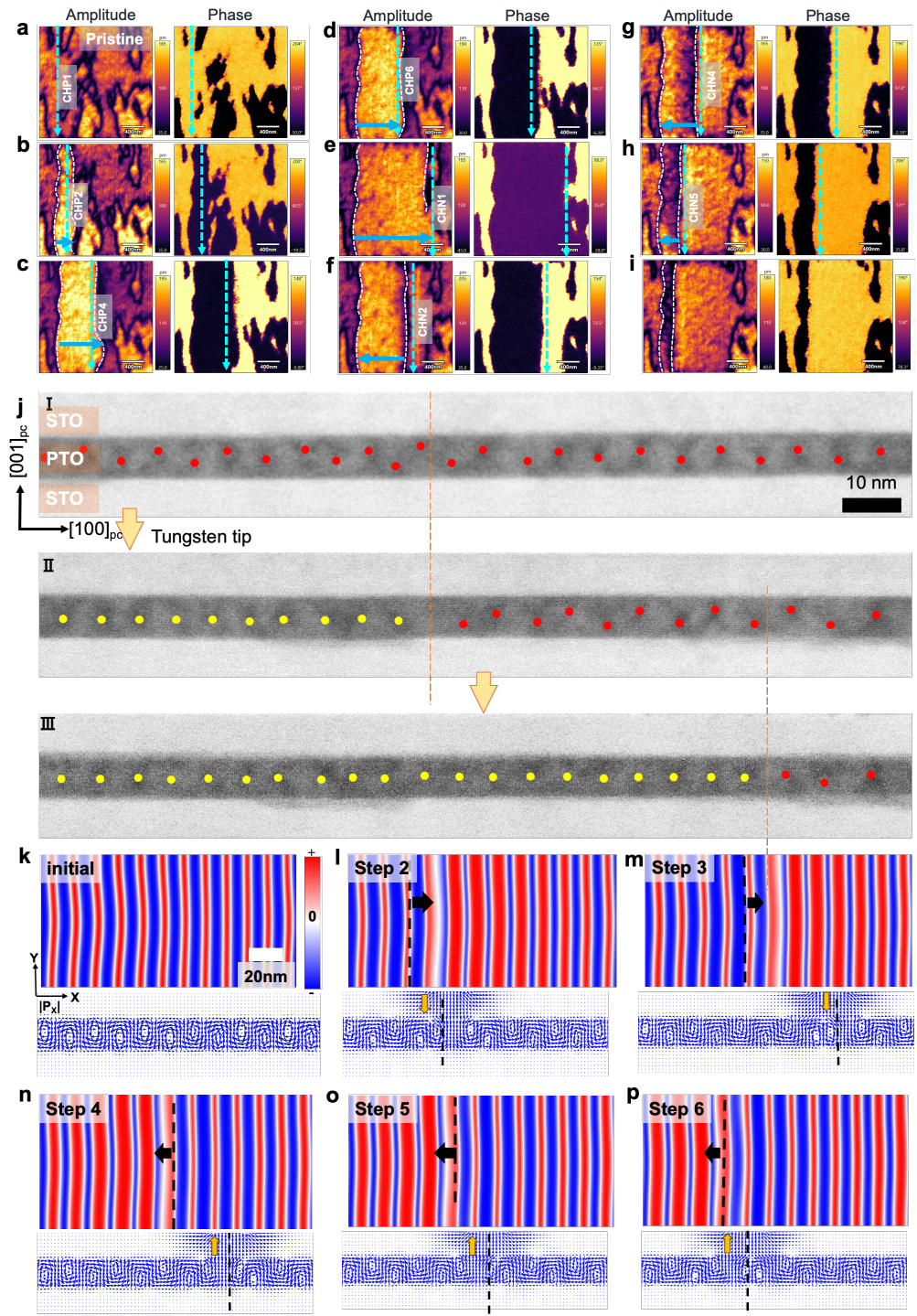

**Figure 2| Harnessing the polar vortex motion in the STO/ PTO/ STO trilayer. a,** Pristine lateral PFM (LPFM) scans (amplitude: left panel; phase: right panel) of the trilayer reveal well-defined pure vortex structures. The cyan arrow indicates the lithographically defined channel where the first positive trailing field (CHP1, +5 V, one minute) was applied. PFM scans were taken after field removal. **b,** Post-field LPFM imaging after CHP1 shows an initial rightward shift of the vortex boundary (highlighted by blue arrow). Sequential trailing fields (CHP2, CHP3, +5 V, one minute each) were applied, further displacing the boundary. **c-e,** PFM scans

after CHP3 reveal a continued rightward shift (highlighted by blue arrow). Additional fields (CHP4, CHP5, +5 V, one minute each) further harness the boundary displacement. After applying CHP6 (+5 V, one minute), the vortex boundary harnessed to total rightward shift of ~1170 nm (e). Details are provided in Supplementary Fig. 3. **f-i,** Panel (e) serves as a reference. The cyan arrow marks the lithographic channel where the first negative trailing field (CHN1; -5V for one minute) was applied. Post-field PFM measurements reveal a leftward vortex boundary shift (blue arrow) (f). Subsequent negative trailing fields (CHN2, CHN3, CHN4, CHN5; -5V, one minute each) were applied sequentially, leading to a cumulative harnessing with leftward displacement of ~890 nm after CHN5 (i). Details are provided in Supplementary Fig. 4. **j,** *In-situ* STEM experiments illustrating vortex boundary movement in the trilayer (I-III). Initially, the vortex cores are arranged in an up-down-up configuration (I, red dots). Application of a +17 V electric field via a tungsten tip (denoted by yellow arrow) aligns the vortex cores into a linear configuration (II, yellow dots), forming a boundary between the up-down arrangement and the linear alignment (orange dashed line). A subsequent +17 V application harnessed the boundary further rightward (III), consistent with the PFM results. After field removal, the vortex cores remain in their displaced positions, demonstrating stability. **k-p,** Phase-field simulations confirm the harnessing of vortex boundary in the trilayer. The initial configuration confirms the presence of vortices (plane view, top panel; cross-sectional view, bottom panel) (k). A positive electric field applied through a designated channel shifts the vortex boundary rightward (l, black dashed line). Reapplication of the positive field causes further rightward displacement (m). Conversely, applying a negative electric field shifts the boundary leftward (n), with successive negative bias fields inducing additional leftward movement (o). Upon field removal, the vortex boundary returns to a configuration similar to the initial state (p).

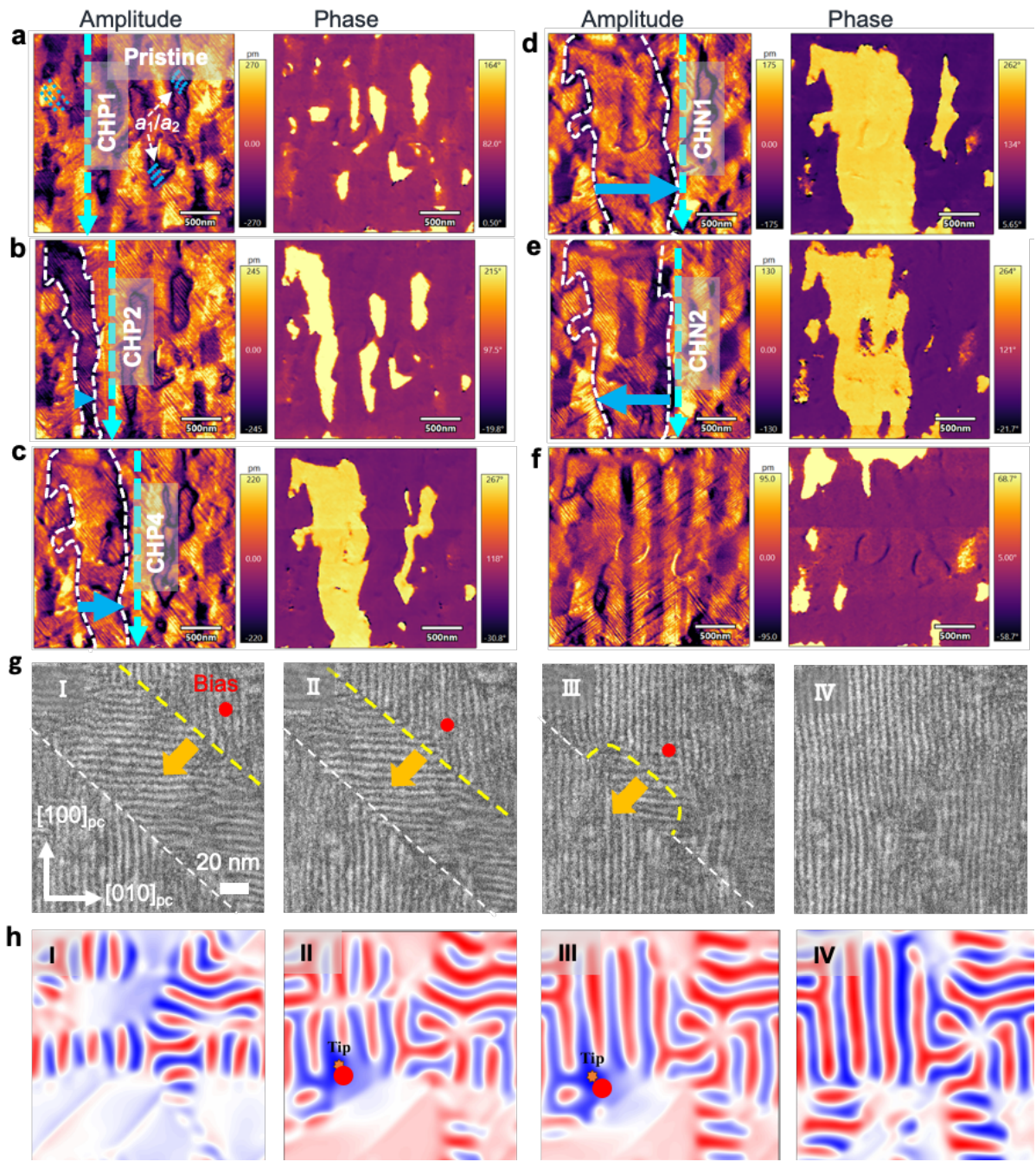

**Figure 3| Harnessing the vortex motion in (PTO$_{16}$/STO$_{12}$)$_{10}$ superlattice. a**, Pristine LPFM scans of the superlattice sample, capturing both amplitude (left panel) and phase (right panel), reveal the presence of vortices (highlighted by white dashed lines) and $a_1/a_2$ superdomains (marked by short dashed lines). The cyan arrow indicates the channel through which the +5V trailing field was applied for one minute, with measurements taken after the field was removed. **b,** PFM scan following the application of the first trailing field through channel (CHP1) for one minute shows the initial movement of the vortex boundary to the right (highlighted by blue

arrow). **c-d,** Successive applications of +5V trailing fields (CHP2-CHP7, each for one minute) progressively harness the vortex boundary further to the right. PFM scans after CHP7 show a maximum harnessing of the vortex in rightward shift of ~1170 nm (blue arrow) (d). This extreme displacement served as the reference point for the subsequent application of a negative trailing field (-5V, CHN1, one minute), which reversed the boundary's movement leftward (details in Supplementary Fig. 4). **e-f,** Additional -5V trailing fields (CHN2, CHN3, one minute each) further shift the boundary to the left, ultimately returning the vortices to their original positions (harnessing of the vortex in leftward shift of ~643 nm), similar to their state in (b). After CHN3, the vortex boundary became disrupted, resembling the pristine vortex structure seen in (a) (details in Supplementary Fig. 5). **g,** *In-situ* STEM images demonstrate vortex boundary manipulation. The pristine area (**I**) shows vortices oriented along the $(100)_{pc}$ and $(010)_{pc}$ directions, with their boundary marked by yellow and white dashed lines. A red point indicates the application of a +20V electric field. The field causes the yellow boundary to move closer to the white dashed line (**II-III**). Further forward movement of the tungsten tip results in the alignment of all vortices along the $(001)_{pc}$ direction (**IV**). **h,** Phase-field simulations validate vortex manipulation in the presence of ferroelectric domains (**I**). Upon applying a bias voltage via the tip, the vortices elongate downward (**II**). As the bias field moves downward, the vortices grow in the field's direction (**III**). The final state after field removal (**IV**) shows stable vortex configurations. These simulations corroborate experimental observations, demonstrating the capability to harness and control vortex behavior.

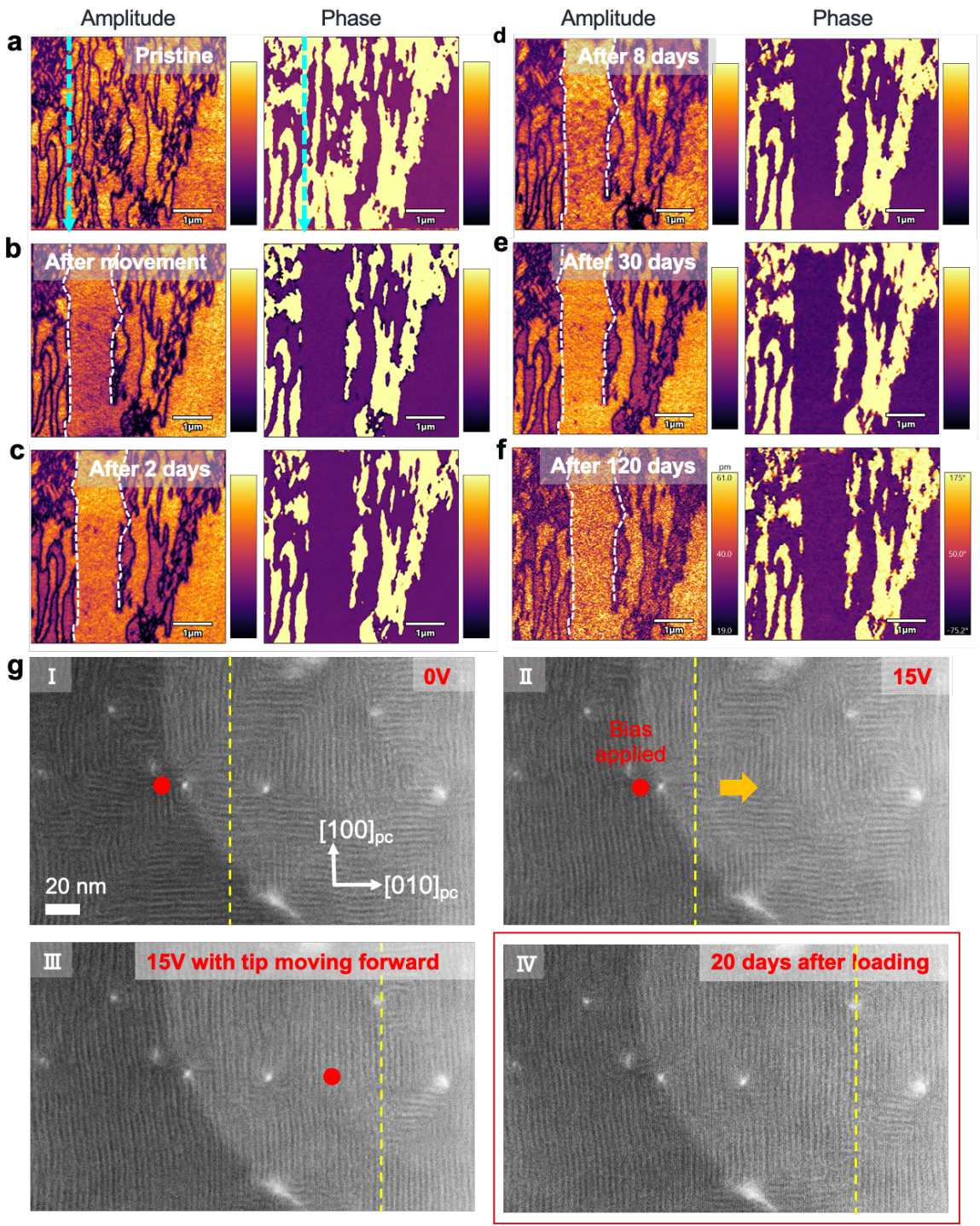

**Figure 4| Stability in harnessing vortex motion. a,** Pristine LPFM images for the PTO/STO/PTO trilayer confirm presence of well-defined vortices. Following the application of a bias electric field through lithographically defined channels, as described in earlier experiments, the vortex boundary shifted from left to right. Subsequent PFM scans conducted at various intervals reveal the long-term stability of the harnessed vortex boundary: **c,** after 2 days; **d,** after 8 days; **e,** after 30 days; and **f,** after 60 days. Each scan demonstrates the sustained

robustness of the harnessed vortex boundary over time. **g,** In-situ STEM images of a plane-view superlattice sample illustrate the movement and stability of vortices. (I) The initial scanned area reveals two vortex orientations, with the boundary marked by a yellow dashed line. A +15V electric field was applied through the tungsten tip at the location indicated by a red dot. In response to the forward movement of the tungsten tip, the vortex boundary shifted rightward, as shown in (II) and (III). After the removal of the bias field, the harnessed vortex boundary remained stable. No observable changes in the vortex boundary were recorded over 20 days (IV), underscoring the robustness and long-term stability of the harnessed vortex boundary.

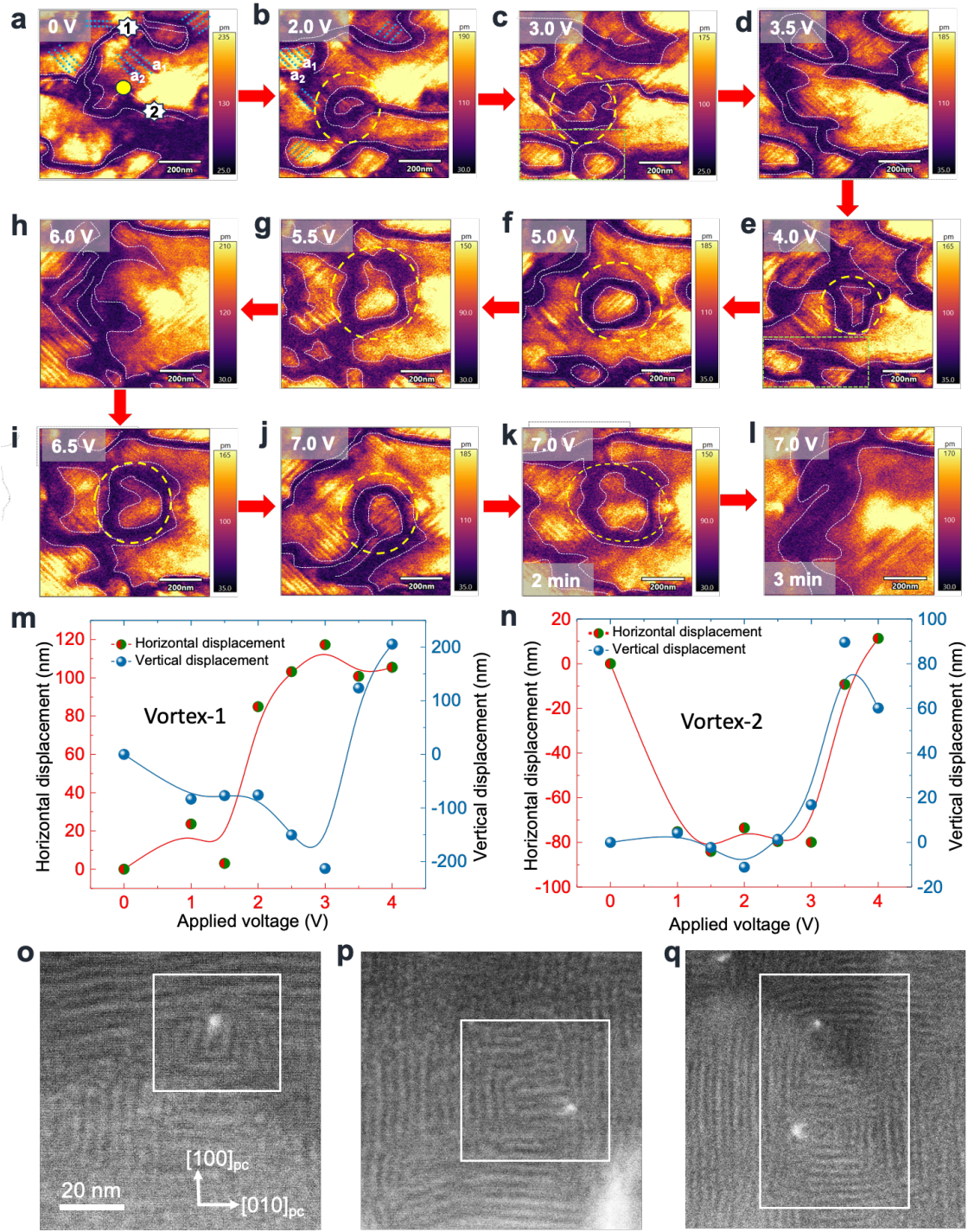

**Figure 5| Stochastic dynamics and vortex ring formation under pulsed fields in the superlattice. a,** Pristine LPFM amplitude image confirms the presence of vortices (dashed white lines) and $a_1/a_2$ superdomains (short blue dashed lines). A yellow circle marks the site of pulsed field application (1–7V, in 0.5V increments, applied for one minute each). Points 1 (vortex 1) and 2 (vortex 2) were tracked for displacement analysis of the vortices. After remove the field, the system was scanned to capture changes in vortex behavior. **b,** A 2V pulsed field applied for one minute at the designated point (yellow marker) induced significant vortex

displacement and the formation of a vortex ring (yellow circle). **c,** After the application of pulsed field of 3 V for one minute, vortex 1 moved away from the bias site, vortex 2 migrated toward it, and a pair of vortex rings formed. **d,** After application of 3.5V pulsed field caused further displacement and reshaping of vortices. The previous vortex ring disappeared, and a new ring emerged in the left corner. Similarly, application of pulsed field of **e,** 4V **f,** 5 V **g,** 5.5 V **h,** 6 V, **i,** 6.5 V, **j,** 7 V for one minute each, **k,** 7 V for two minutes, and **l,** 7 V for three minutes resulted in further morphological transformations and displacement of vortex structures. Vortex rings were observed to form and annihilate in response to the applied biases. The dynamics of vortex 1 and vortex 2, initially marked in (a), were tracked, revealing their displacements in the horizontal (red) and vertical (gray) directions, highlighting the intricate interplay between field magnitude and vortex behavior. **m-n,** Quantification of the displacements of vortex 1 and vortex 2 as a function of the applied pulsed field reveals stochastic behavior in response to varying electric fields. **o-p,** STEM imaging confirms the presence of vortex rings at two distinct locations. **q,** STEM observation further identifies a pair of vortex rings. These STEM findings corroborate the PFM-based observations of vortex rings and paired vortex rings.